\documentclass[prl,aps,10pt,superscriptaddress,noshowpacs,preprintnumbers,twocolumn]{revtex4-1}

\usepackage{float}
\usepackage{epsfig,amsopn,amsthm}
\usepackage{graphicx,epstopdf}
\usepackage{sidecap}
\usepackage{subfiles}
\usepackage{amsmath,amssymb}
\usepackage{enumerate}
\usepackage{dsfont,verbatim,varwidth}
\usepackage[usenames, dvipsnames]{color}

\begin{document}

\title{Emergence of spin-active channels at a quantum Hall interface}

\author{Amartya Saha}
\affiliation{Department of Physics and Astronomy, University of Kentucky, Lexington KY 40506-0055, USA}
\author{Suman Jyoti De}
\affiliation{Harish-Chandra Research Institute, HBNI, Chhatnag Road,  Jhunsi, Allahabad 211019, India}
\author{Sumathi Rao}
\affiliation{Harish-Chandra Research Institute, HBNI, Chhatnag Road, Jhunsi, Allahabad 211019, India}
\author{Yuval Gefen}
\affiliation{Department of Condensed Matter Physics, Weizmann Institute, 76100 Rehovot, Israel}
\author{Ganpathy Murthy}
\affiliation{Department of Physics and Astronomy, University of Kentucky, Lexington KY 40506-0055, USA}

\date{\today}

\begin{abstract}
We study the ground state of a system with an interface between
$\nu=4$ and $\nu=3$ in the quantum Hall  regime. Far from the
interface, for a range of interaction strengths,  the $\nu=3$ region is
fully polarized but $\nu=4$ region is locally a singlet. Upon varying
the strength of the interactions and the width of the interface, the
system chooses one of two distinct edge/interface phases. In phase A,
stabilized for wide interfaces, spin is a good quantum number, and
there are no gapless long-wavelength spin fluctuations. In phase B,
stabilized for narrow interfaces, spin symmetry is spontaneously
broken at the Hartree-Fock level. Going beyond Hartree-Fock, we argue that
phase B is distinguished by the emergence of gapless long-wavelength
spin excitations bound to the interface, which can, in principle, be
detected by a measurement of the relaxation time $T_2$ in nuclear
magnetic resonance.
\end{abstract}

\maketitle

\noindent \emph{Introduction: }In the integer quantum Hall effect (IQHE),
\cite{Klitzing} a two-dimensional electron gas (2DEG) subjected to a
strong perpendicular magnetic field displays a Hall conductivity
quantized in integral units of $\frac{e^2}{h}$ at low
temperatures. These systems are the simplest examples
of topological insulators (TIs)\cite{TIrev}. Their bulk is insulating,
and the underlying band topology manifests itself in chiral
current-carrying edge states which are protected against
localization. The topological nature of the bulk state dictates the
charge Hall and thermal Hall conductances. In addition, because the
kinetic energy is quantized into degenerate Landau levels (LLs),
partially filled LLs host the strongest possible electron
correlations, leading to quantum Hall
ferromagnetism\cite{QHferromag1,KunYang} and the fractional quantum
Hall effects\cite{Tsui}.

Edges play a central role in the QHE, and it has long been realized
that within the topological constraints imposed by the bulk, a variety
of reconstructed edge phases are possible. Much theoretical work
exists on edge reconstructions, with most reconstructions being driven
by electrostatic considerations: The ``desire'' of the electron fluid
to perfectly neutralize the positive background competing with the
``desire'' to form incompressible droplets. In the simplest
reconstructions spin plays no
role\cite{Chklovskii,Chamon,Dempsey,Meir,JWang,MacDonald2/3}.

Generically, at the edges of quantum Hall ferromagnets, states with
broken spin and/or edge translation symmetry are known to occur in the
Hartree-Fock (HF) approximation. \cite{Sondhi,FrancoandBrey,Oakninedge,Khanna}

It is clear from previous work that edge reconstructions can generate
counterpropagating pairs of chiral charge modes.  Going beyond
previous work, one can ask whether exchange can lead to the emergence
of a pair of chiral, neutral, spin-active edge modes. Since spin is
involved, at least one of the two bulk quantum Hall states must be a
QH ferromagnet.

Motivated by these considerations, we investigate an interface between
a $\nu=4$ singlet region and a fully polarized $\nu=3$ region in the
Hartree-Fock (HF) approximation. In the following, we will use the
words edge and interface interchangably. Our tuning parameters are the
width of the interface measured in units of magnetic length (${\tilde
  w}=w/\ell$), where the background charge is assumed to vary smoothly
between $\nu=4$ and $\nu=3$, and the strength of the Coulomb
interaction relative the the cyclotron energy (${\tilde
  E}_c=\frac{e^2}{\varepsilon\ell\hbar\omega_c}$). We find two robust
phases: For large ${\tilde w}$ we find phase A: all HF single-particle
levels are spin polarized and three of them cross the Fermi energy, as
required by the total $S_z=0$ in the $\nu=4$ bulk and the total
$S_z=3/2$ in the $\nu=3$. For smaller ${\tilde w}$ we find phase B,
with spontaneously broken $U(1)$ spin-rotation symmetry, and a single
HF level crossing the Fermi energy appears.

Phase A is conventional, with a pair of counterpropagating, spin-resolved  chiral
charge modes in addition to the one chiral charge mode required by
topology. Phase B, as we will argue in the discussion, manifests a
pair of chiral, counterpropagating spin-active neutral modes bound to
the interface, in addition to the required charged chiral. Any probe
sensitive to gapless long-wavelength spin excitations, such at nuclear
magnetic resonance (NMR), will be able to distinguish the two phases.

In the following, we will set up the problem, explain our computation
briefly, and describe the two phases in HF. We address the important
issue of fluctuations beyond the HF approximation in the discussion,
before addressing potential experimental signatures. Details of the
robustness of the two phases with respect to the Zeeman coupling ($\tilde E_Z=\frac{E_Z}{\hbar \omega_c}$), the
screening of the interaction, the number of Landau levels kept in our
calculation, and other details,  are relegated to the supplemental
material (SM) \cite{supp}.

\noindent \emph{Edge between $\nu=4$ and $\nu=3$ quantum Hall states}:
The geometry of the interface between the $\nu=4$ and $\nu=3$ QH
systems is shown in Fig.\ref{setup}.  In the non-interacting limit,
the $\nu=4$ bulk will have the Landau levels (LLs) $|0\uparrow,
0\downarrow, 1\uparrow, 1\downarrow>$ occupied, while the $\nu=3$ bulk
has the LLs $|0\uparrow, 0\downarrow,1\uparrow>$ occupied.  In this
case, at the edge between the two, we expect the $1\downarrow$ LL to
smoothly cross the chemical potential $\mu$ from below as one
moves rightwards (from $\nu=4$ to $\nu=3$), leading to  a single chiral
charged edge mode with $\downarrow$-spin.

\begin{figure}[H]
\centering
\includegraphics[width=0.8\columnwidth]{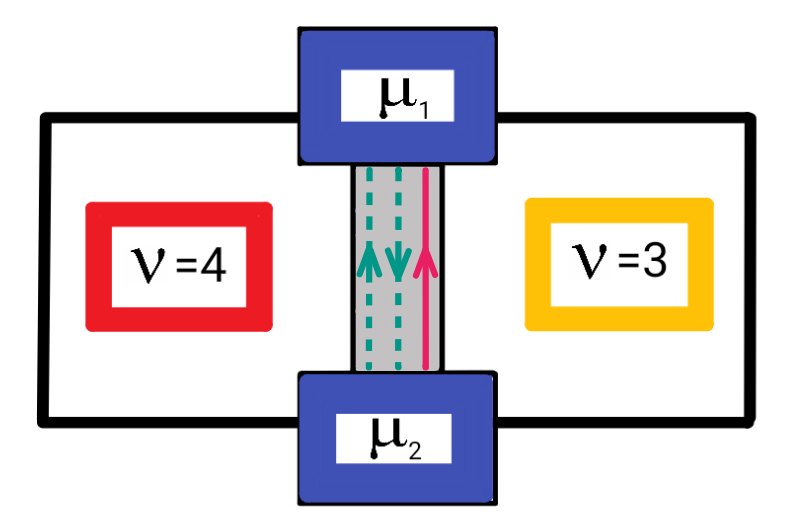}
\caption{A schematic diagram of our setup with an interface between
  bulk $\nu=4$ and $\nu=3$ IQHE states. The solid line (red online) is
  a downstream chiral charged mode required by topology.  The pair of
  dashed lines (green online) depict either spin-resolved charged
  chiral modes (phase A) or gapless spin-active chiral modes (phase B). }
\label{setup}
\end{figure}

 As interactions grow stronger we expect a greater tendency towards
 spin polarization (QH ferromagnetism).  However, the $\nu=3$ and
 $\nu=4$ states do not get polarized at the same value of
 $\tilde{E}_c$. There is a range of $\tilde{E}_c$ where the $\nu=4$
 bulk remains unpolarized, while the $\nu=3$ bulk is fully
 polarized. Now it is no longer obvious how many $\mu$ crossings, and
 hence chiral modes, there should be: The result will depend on the
 details of the interface.  Our goal is to study the possible edge
 phases that can exist as our tuning parameters
 $\tilde{w},\tilde{E}_c$ are varied.

The  bulk  Hamiltonian for a quantum Hall system  is
\begin{eqnarray}
        H &=&\hbar\omega_c\sum_{n,k,s}c^\dagger_{nks}c_{nks} +\frac{E_Z}{2}\sum_{n,k}(c^\dagger_{nk\downarrow}c_{nk\downarrow}-c^\dagger_{nk\uparrow}c_{nk\uparrow}) \nonumber\\
        &+&\frac{1}{2\pi A}  \sum_{{\bf q}} v({\bf q})   (\rho_b(-{\bf q})-\rho_e(-{\bf q}))\nonumber\\
        &&(\rho_b(-{\bf q})-\rho_e(-{\bf q})) .
\end{eqnarray}
Using $n$ for the Landau level index and $k$ for the guiding center
index (defined below), the electron density operator is
$\rho_e(x,y)=\sum\limits_{s}\Psi_s^\dagger(x,y)\Psi_s(x,y)$, where the
electron field operator is
$\Psi_{s}(x,y)=\sum\limits_{n,k}\Phi_{nk}(x,y) c_{nks}$, with
$c_{nks}$ being canonical fermion operators.  $v({\bf q})$ and
$\rho_e({\bf q})$ are the Fourier transforms of the long-ranged
screened Coulomb potential $v({\bf {r}}-{\bf{r'}})$ and $\rho_e(x,y)$
respectively.  We model the background charge density $\rho_b$ as
changing linearly from $4\rho_0$ to $3\rho_0$ ($\rho_0$ is the density
of a single filled Landau level) over a distance $\tilde w$ in the
${\hat y}$ direction as shown in Fig.\ref{background}.  Note that the
background charge density preserves translation invariance in the x
direction.  Thus, $\tilde w$ serves as the tuning parameter which
controls the softness of the background potential near the interface.
As in real samples, the Zeeman coupling $\tilde E_Z>0$ (but $\tilde
E_Z\ll \tilde E_c$ and $\hbar\omega_c$)).  It follows that the spin
symmetry of the Hamiltonian is $U(1)$.

\begin{figure}[H]
\centering
\includegraphics[width=0.85\columnwidth]{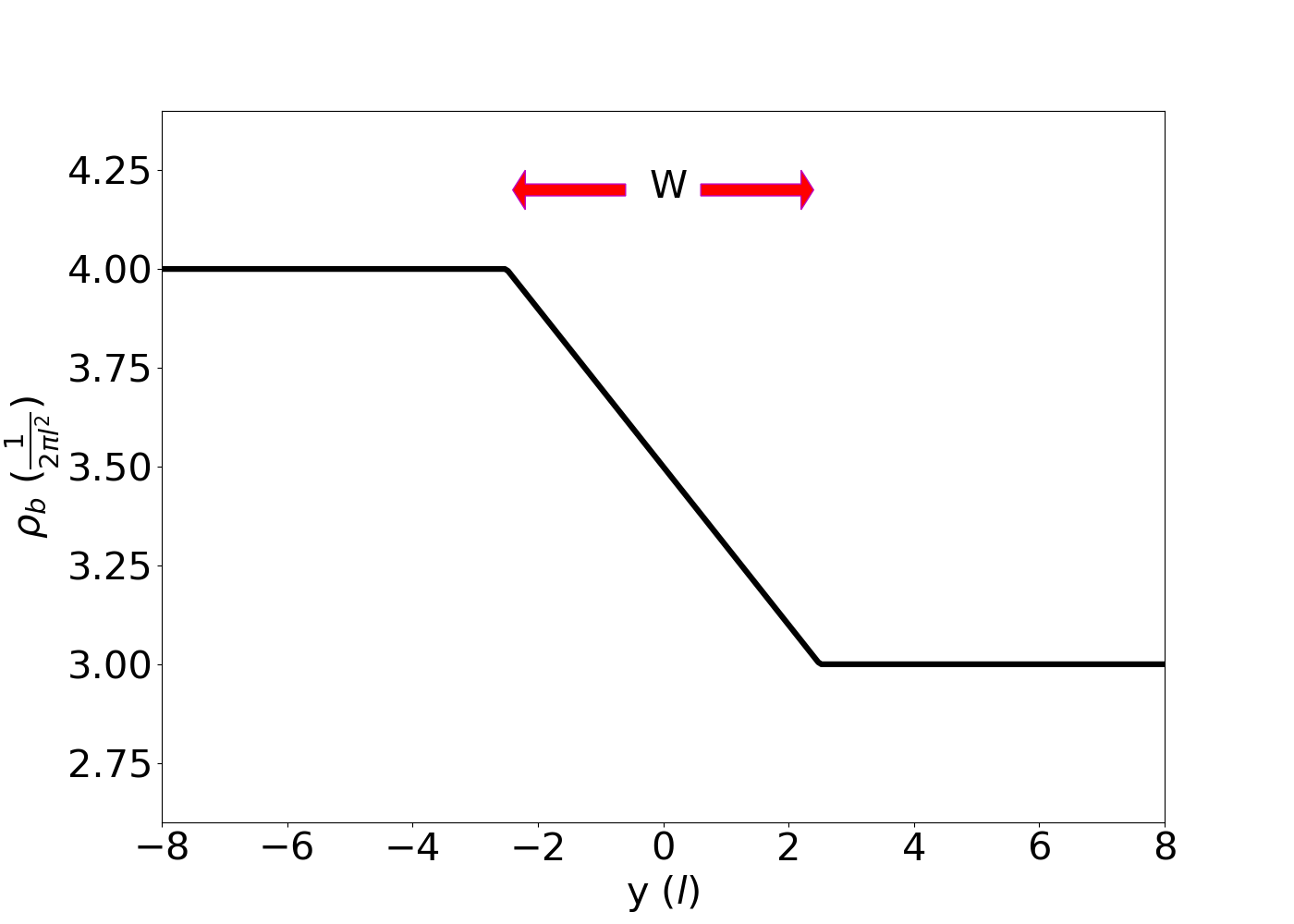}
\caption{Dependence of the background charge density on ${\hat y}$. The charge density
is uniform in the ${\hat x}$ direction.}
\label{background}
\end{figure}

For the unscreened Coulomb interaction, at $\tilde E_Z=0$ in the HF approximation, for
$2.52<{\tilde E}_c<2.90$ the bulk $\nu=4$ ground state
($0\uparrow,0\downarrow,1\uparrow,1\downarrow$ occupied) is unpolarized
and the bulk $\nu=3$ ground state
($0\uparrow,1\uparrow,2\uparrow$ occupied) is fully polarized. As $E_z$ increases the
range of ${\tilde E}_c$ changes. 

We work in the Landau gauge $\vec{A}=-B_0y \hat{x}$ with the magnetic
field pointing in the positive ${\hat z}$-direction. The magnetic
length $l=\sqrt{\frac{\hbar}{eB_0}}$. The Hamiltonian has translation
invariance along $x$ (even with the interface potential). The one-body
wavefunctions are
\begin{equation}
    \Phi_{n,k}(x,y)=\frac{e^{ikx}e^{-\frac{(y-kl^2)^2}{2l^2}}}{\sqrt{L_x n!2^n\sqrt{\pi}l}} H_n(\frac{y-kl^2}{l}).
\end{equation}

The $ x$ coordinate (along the edge) has periodic boundary
conditions to discretize $k$, which defines the guiding centre
position $Y(k)=kl^2$. The interface is centred at $y=0$
with $\nu=4$ as the bulk ground state for $y<0$ and $\nu=3$ as the
bulk ground state for $y>0$.  We work in spin-unrestricted HF
theory, and look for solutions that preserve the translation
invariance in $x$ of the Hamiltonian, implying that $k$ is a good
single-particle quantum number in HF. Since we allow for Landau level
and spin-mixing, we work with a total of 8 basis states for each value
of the guiding centre $k$,
consisting of 4 Landau levels, each with $\uparrow$ and $\downarrow$
spins.  The translational invariant ground states of the theory are
defined in terms of the matrix $\Delta_{ns,n's'}(k)=\langle c^\dagger
_{n's'k} c_{nsk}\rangle$, which is obtained self-consistently by
diagonalizing the effective one-body Hartree-Fock Hamiltonian. The
chemical potential $\mu$ is chosen to maintain overall charge
neutrality. We use a screened Coulomb potential of the form
$v(q)=\frac{2\pi E_c}{q+q_0}$, where $q_0$ , the screening parameter,
is chosen to be $q_0=0.01$. Using this method we obtain the phase
diagram in the parameters $\tilde{w},\ \tilde{E}_c$.

The SM \cite{supp} contains the details of the HF procedure, and an
analysis of the stability of our phase diagram with respect to
variations in the screening wavevector $q_0$, the Zeeman coupling
$E_z$, and the number of Landau levels that we keep in our calculation.

\noindent \emph{Phase diagram in the Hartree-Fock approximation:}
There are two distinct edge phases, as shown in Fig.\ref{phasediag},
separated by a first-order phase transition. In phase A, which is
stabilized for very smooth edges, there are three $\mu$-crossings of
single-particle levels, each spin-resolved. In phase B, stabilized for
relatively sharp edges, there is only a single self-consistent energy
level that crosses $\mu$. In addition, the HF state of phase B shows a
spontaneous breaking of the $U(1)$ spin symmetry.

\begin{figure}[H]
\centering
\includegraphics[width=0.85\columnwidth]{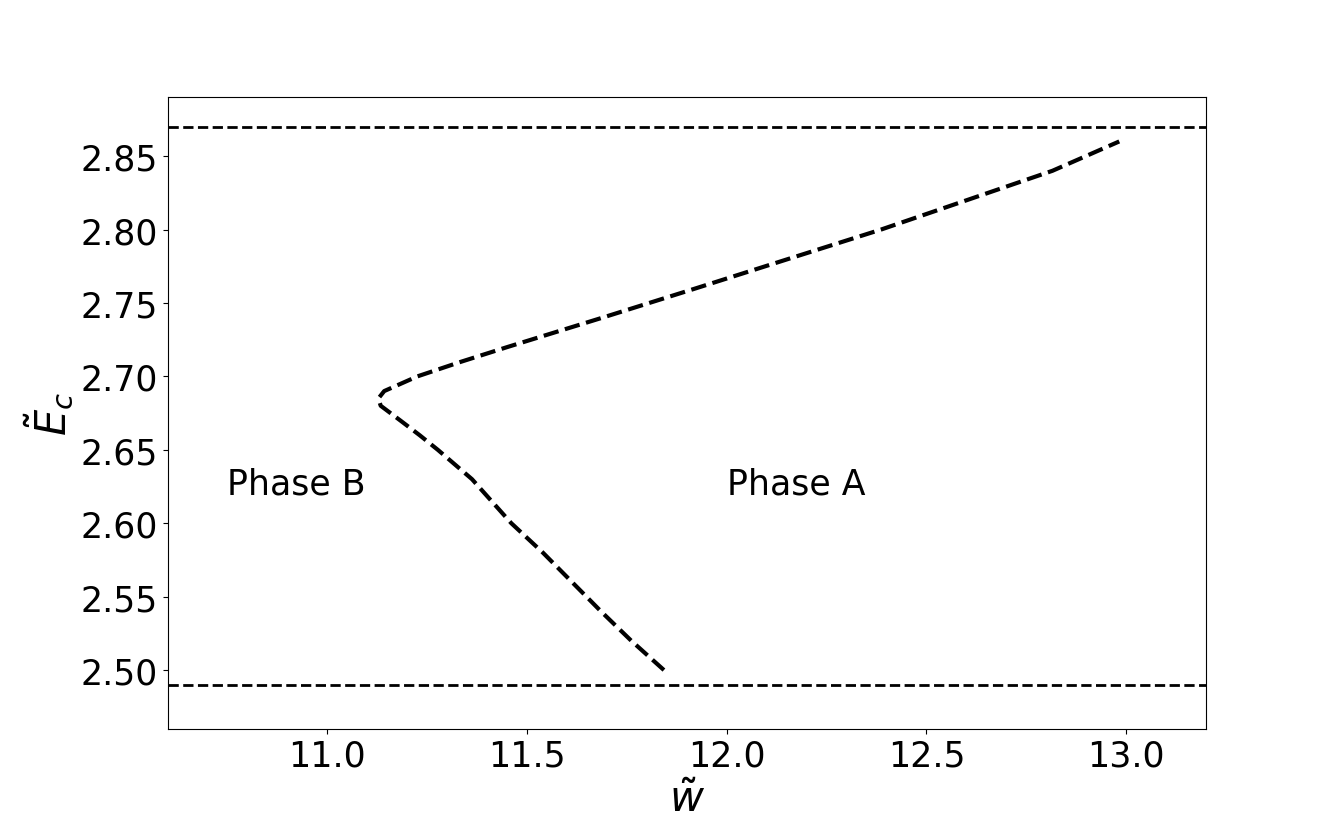}
\caption{Phase diagram in the parameter space $\tilde{E}_c$ and
  $\tilde{w}$ at $\tilde E_Z=0.03 $. At this value of $\tilde E_Z$ the $\nu=4$ 
  bulk state is a singlet and $\nu=3$ fully polarized for $2.49<{\tilde E}_c<2.87$. For values of ${\tilde
    E}_c<2.7$ Landau level mixing is not a significant effect, and the
  spin-stiffness increases with ${\tilde E}_c$. This raises the cost
  of phase B over phase A, leading to the phase boundary moving
  towards smaller ${\tilde w}$.  For ${\tilde E}_c>2.7$ Landau level
  mixing decreases the spin-stiffness, thereby favoring phase B. The
  transition is first-order in HF.}
\label{phasediag}
\end{figure}

The main features of the phase diagram result from the competition
between (i) the interface potential, controlled by the width $\tilde w$ of
the interface region, (ii) the electrostatic repulsion represented by
the Hartree term and, (iii) the spin stiffness governed by the Fock
term. All three are controlled by the Coulomb interaction. For large
values of $\tilde w$, it is energetically favourable for the system to
approximately neutralize the background potential by creating an extra
pair of counter-propagating edge modes, spreading the electron density
over a larger region. In this phase, the spins of the chiral modes
(assuming one associated with each single-particle $\mu$-crossing)
remain well-defined. For smaller values of $\tilde w$, it becomes
energetically favourable to have a single HF level crossing $\mu$. The
requirement that the spin polarization at each $k$ change by
$\frac{3\hbar}{2}$ in going from $\nu=4$ to $\nu=3$ necessitates a
spin rotation (and Landau level index rotation) of all single-particle
HF levels, especially occupied ones.

\begin{figure}[H]
\centering
\includegraphics[width=0.8\columnwidth]{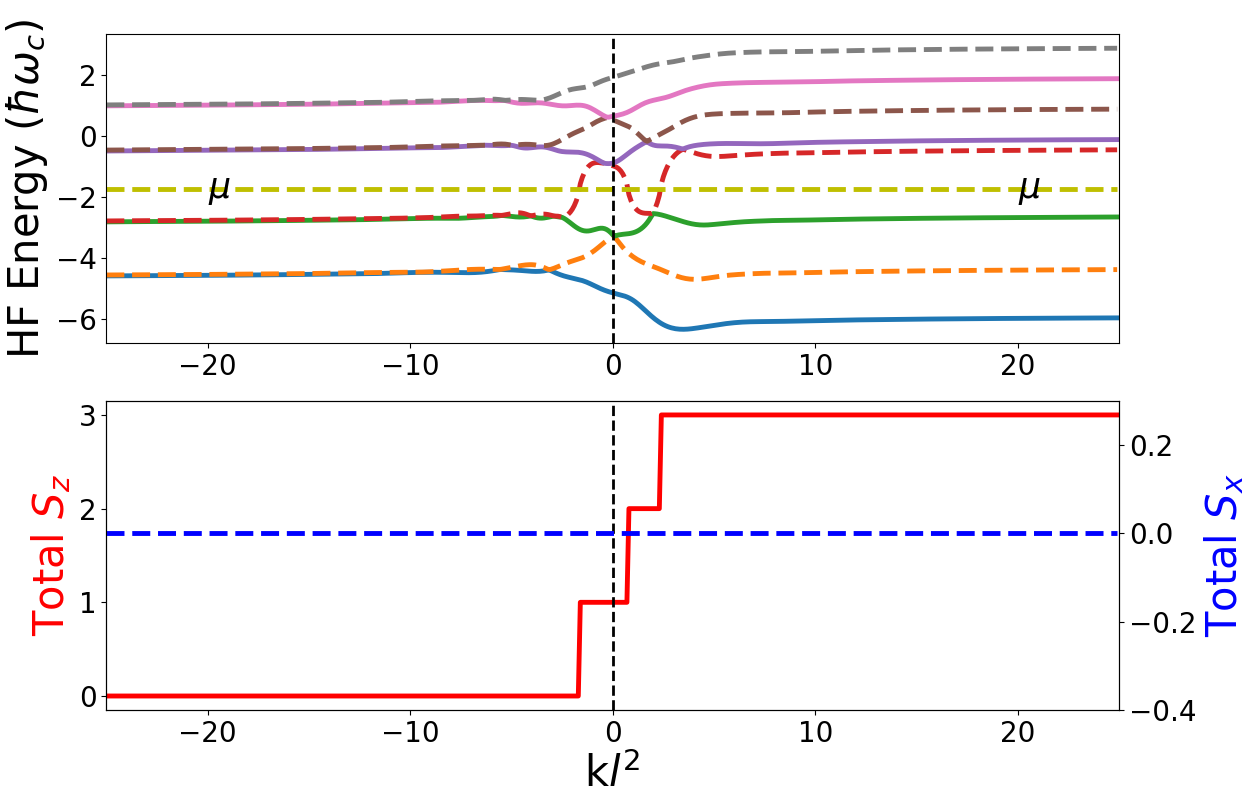}
\caption{Phase A in the HF approximation: The upper panel shows the single-particle energy
  dispersion and the lower panel shows the total $S_z$ and $S_x$
  values (in units of $\frac{\hbar}{2}$) as a function of the guiding
  center position. The parameter values are $\tilde{E}_c=2.52$, $\tilde{w}=13.0$ and $\tilde E_Z=0.03 $ }
\label{phaseA}
\end{figure}

\begin{figure}[H]
\centering
\includegraphics[width=0.8\columnwidth]{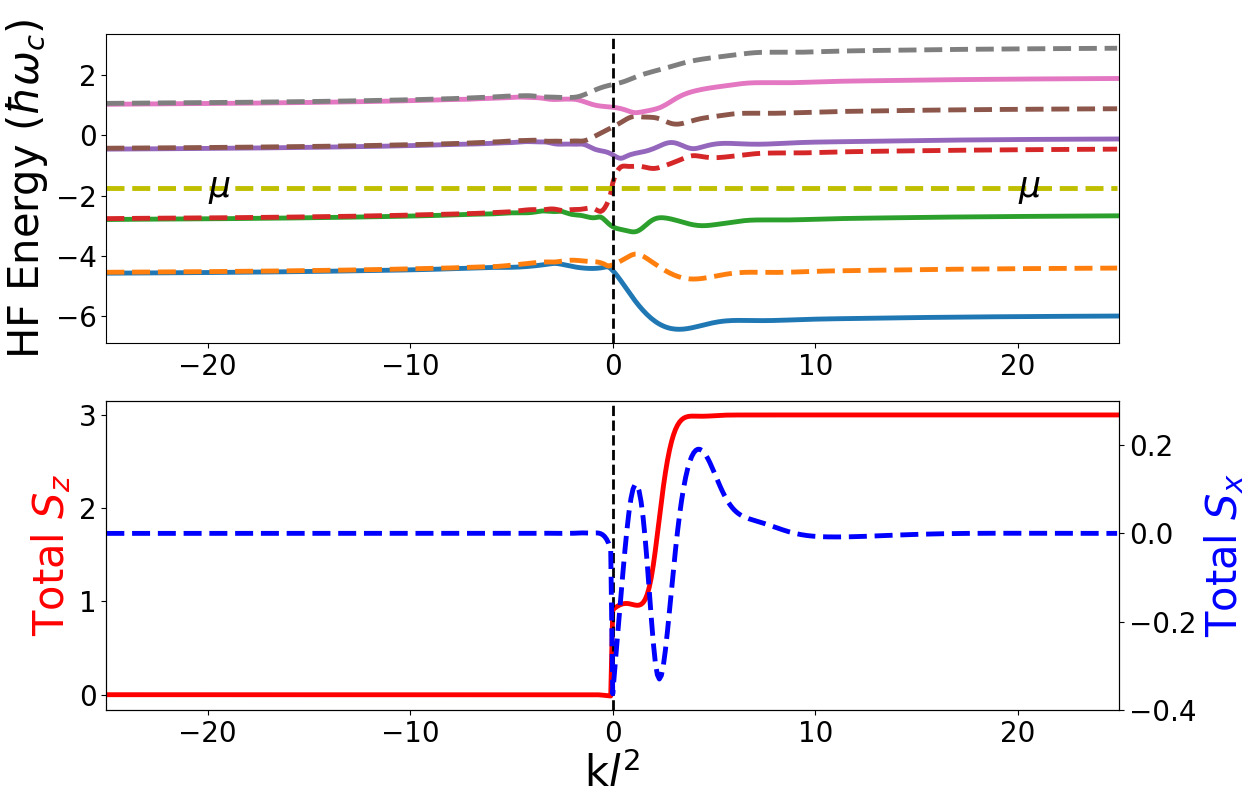}
\caption{Phase B in the HF approximation: The upper panel shows the single-particle energy
  levels and the lower panel shows the total $S_z$ and $S_x$ values
  (in units of $\frac{\hbar}{2}$) as a function of the guiding center position.
  The parameter values are $\tilde E_c=2.52$, $\tilde w=10.0$ and $\tilde E_Z=0.03 $}
\label{phaseB}
\end{figure}

Let us examine more closely what we can infer from the HF
solution. For this paragraph only, we will make the naive assumption
that each single-particle crossing of $\mu$ represents a chiral mode.
The single particle energy levels and the spin components of the
levels are plotted in Figs. \ref{phaseA} and \ref{phaseB}.  From the
energy dispersions in Fig.\ref{phaseA} we see that two of the modes in
phase A are downstream and one is upstream. The spin of these modes
can also be explicitly identified as follows.  Moving rightwards from
large negative $y$, first the 1$\downarrow$ level from the $\nu=4$
region smoothly crosses $\mu$ from below, implying a downstream chiral
mode with $\downarrow$ spin. Next the 2$\uparrow$ level crosses $\mu$
from above, producing an upstream chiral mode with $\uparrow$
spin. Finally the 0$\downarrow$ level crosses $\mu$ from below,
producing another downstream chiral mode with $\downarrow$ spin. Note
that the spin at the interface changes by three units (each unit is
$\frac{\hbar}{2}$) as is required since the bulk state for $y<0$ is
the unpolarized $\nu=4$ quantum Hall state and the bulk state for
$y>0$ is the fully polarized $\nu=3$ quantum Hall state. Note also
that the average value of $S_x$ remains zero confirming that the
chiral levels have well-defined spins. Phase B, on the other hand, has
only one chiral downstream mode. Here again, the spin at the interface
does change by three units as is required, but the average value of
$S_x$ is non-zero here and there is spin rotation at the interface.

\noindent \emph{Discussion of fluctuations beyond HF}: HF is known to
broadly overpredict order, due to its neglect of quantum
fluctuations. Thus, HF can be taken as reliable for single-particle
spectra, but should be supplemented by reasoning based on effective
theories when questions of spontaneously broken symmetry and
collective modes arise. We will proceed in three steps. (i) We
identify the correct effective theory based on symmetries and
dimensionality. (ii) We match the HF phases to those of the effective
theory by considering a mean-field limit of the effective
theory. (iii) We look at quantum fluctuations beyond mean-field in the
effective theory, and the implied consequences for physical
observables in our system.

The $SU(2)$ spin symmetry of our electronic Hamiltonian is broken to
$U(1)$ spin rotation around total $\vec{B}$, in the presence of Zeeman
coupling, and the edge is a quasi-1D system. Thus, the relevant
effective theory in the spin sector is the XXZ model in a Zeeman field
in 1D\cite{sachdev2011,giamarchi2004quantum}.  $H_{xxz}=-J\sum S_x(n)S_x(n+1)+S_y(n)S_y(n+1)+
\Delta S_z(n)S_z(n+1)-E_z\sum S_z(n)$.

There is some complicated mapping between our tuning parameters
$\tilde{w}, \tilde{E}_c$ and the XXZ parameters $J,\ \Delta$.To match
the phases in HF with those of $H_{xxz}$, we take the classical limit
of the latter. For $\Delta<1$ and $\ E_z<4J(1-\Delta)$,  the XXZ model
spontaneously breaks the $U(1)$ symmetry classically, while for
$\Delta>1$,  it does not. We conclude that $\Delta<1$ in phase B of HF,
while $\Delta>1$ in phase A.

We finally come to the important issue of quantum fluctuations beyond
HF. The Mermin-Wagner theorem\cite{MerminWagner} ensures that a
continuous symmetry cannot be spontaneously broken in 1D, even at zero
temperature.  Hence the spontaneous breaking of the $U(1)$
spin-rotation symmetry seen in HF (and the classical limit of
$H_{xxz}$) will not survive quantum fluctuations. {\it However,
  $H_{xxz}$ still has two distinct phases}. The distinction between
the phases lies in the presence of gapless long-wavelength spin
excitations for $\Delta<1$, while they are absent for $\Delta>1$.

The physical consequences for our model system, the $\nu=4$ to $\nu=3$
interface, are striking. Phase B will have, in addition to the charged
chiral edge mode predicted in HF, a pair of gapless, chiral,
counterpropagating spin modes bound to the interface. Phase A, on the
other hand, simply has three well-separated charged spin-resolved
chiral modes (two downstream and one upstream). There are no gapless
long wavelength spin-flip excitations in phase A.

The classical analysis of $H_{xxz}$ shows that the system can undergo
the $B\rightarrow A$ transition even for $\Delta<1$ upon increasing
$E_z$. This agrees with the HF analysis (see SM\cite{supp}) in which
increasing $E_z$ favors phase A over phase B. It should thus be
possible to drive the $B\rightarrow A$ transition in a given sample by
applying an in-plane field.

Let us now examine experimental signatures that distinguish phases A
and B. Any probe that couples to low-energy long-wavelength spin
fluctuations can be used to tell the phases apart. One such probe is
NMR. The nuclear spin moments couple to the external field via their
own Zeeman term, and to the electronic spins via the hyperfine
interaction. The total electronic spin polarization is measured by the
Knight shift \cite{slichter1996principles} of the frequencies of NMR
resonance lines. This method has been used to measure the total spin
polarization of QH ferromagnets
\cite{BarrettOPNMR1,BarrettOPNMR2,NMR1}. The macroscopic nuclear spin
moment relaxes in two ways, firstly via the inhomogeneous distribution
of local effective magnetic fields (with relaxation time $T_1$), and
secondly via true energy relaxation by emitting and absorbing
low-energy electronic spin degrees of freedom (the relaxation time
$T_2$). Clearly, $T_2$ is the relevant quantity to detect the presence
of absence of gapless electronic spin excitations. A transition from A
to B will lead to a dramatic increase of the energy relaxation rate of
nuclear spins, and thus a decrease of $T_2$. One complication in our
system is that only the nuclear spins near the interface will couple
to the gapless chiral spin modes, so a local measurement of $T_2$ will
be necessary. Some progress has been made in this direction
recently\cite{NMR2,NMR3}.

We leave several important questions for future analysis. (i) Are
there phases besides A and B in a physically realistic model? It seems
theoretically possible that in phase B, quantum fluctuations could gap
out the spin excitations while leaving the chiral charged edge mode as
the sole survivor. Such a phase (call it $B^*$) would be distinct from A
because upstream modes (measurable in two-terminal interface
charge/thermal conductance) are present in A but absent in $B^*$. (ii)
What is the order of the $A\rightarrow B$ transition? The XXZ suggests
a $2^{nd}$ order transition while HF implies $1^{st}$ order. (iii) Can the
gapless chiral spin modes in phase B carry charge? It seems possible 
that they can, based on the spin-charge relation in QH ferromagnets.

\acknowledgements

We would like to thank Udit Khanna for many illuminating discussions. We would also like to
thank the International Center for Theoretical Sciences, Bangalore,
for its hospitality and support during the workshop ``Novel Phases of
Quantum Matter'' (Code: ICTS/Topmatter2019/12). AS and GM would like
to thank the US-Israel Binational Science Foundation for its support
via grant no. 2016130. SR and GM would like to thank the VAJRA
scheme of SERB, India for its support. 
YG was also supported by CRC 183 (project C01) of the DFG, 
the Minerva  Foundation,  DFG  Grant  No.  RO  2247/8-1,  
DFG Grant No. MI 658/10-1 and the GIF Grant No. I-1505-303.10/2019.
We would also like to thank the University of Kentucky Center for Computational
Sciences and Information Technology Services Research Computing 
for their support and use of the Lipscomb Compute Cluster 
and associated research computing resources.

\onecolumngrid
\clearpage

\renewcommand{\thefigure}{S\arabic{figure}}
\setcounter{figure}{0}
\renewcommand{\theequation}{S\arabic{equation}}
\setcounter{equation}{0}
\renewcommand\thesection{S\arabic{section}}
\setcounter{section}{0}

\title{Supplemental material for \\ Emergence of spin-active channels at a quantum hall interface}

\author{Amartya Saha}
\affiliation{Department of Physics and Astronomy, University of Kentucky, Lexington KY 40506-0055, USA}
\author{Suman Jyoti De}
\affiliation{Harish-Chandra Research Institute, HBNI, Chhatnag Road, Jhunsi, Allahabad 211019, India}
\author{Sumathi Rao}
\affiliation{Harish-Chandra Research Institute, HBNI, Chhatnag Road, Jhunsi, Allahabad 211019, India}
\author{Yuval Gefen}
\affiliation{Department of Condensed Matter Physics, Weizmann Institute, 76100 Rehovot, Israel}
\author{Ganpathy Murthy}
\affiliation{Department of Physics and Astronomy, University of Kentucky, Lexington KY 40506-0055, USA}

\maketitle

In this set of supplemental materials, we provide the details of our
theoretical calculations and, importantly, the checks that we have
made regarding the robustness of the phases and the phase diagram with
respect to various deformations of the theory.  In Section I, we
establish our notation and briefly recapitulate the Hartree-Fock (HF)
method. Next, in Section II we study how the phase diagram changes
when we include different numbers of Landau levels in our
calculation. Here, we present the phase diagram with three Landau
levels and contrast it to the phase diagram with four Landau levels
(shown in the main text).  In Section III, we study how the
phase diagram changes when we change the Zeeman energy and see how
phase A expands (in most of the parameter space) at the expense of
phase B when we increase the Zeeman energy. We also describe the
various checks that we made of the stability of the phase diagram
to a change in the screening length, and to changes in the ratio of
the Hartree and exchange terms.

\section{I.\,\,\,\,\,\,    The   basic setup and notations}

The notation used here essentially follows the notation given in the
supplemental material of an earlier paper\cite{Khanna2017} co-authored
by some of the present authors.  As discussed in the main paper, we
are studying the ground state of  a system with an interface between a
$\nu=3$ quantum Hall region and a $\nu=4$ quantum Hall region.  We study
the interacting Hamiltonian by using the self-consistent Hartree-Fock
approximation.  We first decompose the interaction term in the Hamiltonian
into Hartree ($V_H$) and Fock ($V_F$) terms by assuming  averages of
the form
\begin{align}
\langle c^\dagger_{n_1k_1s_1} c_{n_2k_2s_2}\rangle = \delta_{k_1,k_2} \Delta_{n_1s_1;n_2s_2}(k_1).
\end{align}
We assume that the state has translation invariance along
x-direction (along the edge), which leads to the guiding center index
$k$ being a conserved quantity in HF.

In the translation invariant bulk there is no mixing between Landau
levels in HF. Further, when the Zeeman term is nonzero,  the spin of the
single-particle bulk HF states are good quantum numbers, implying that $\Delta$
becomes diagonal in both Landau level and spin indices and is given by
the single-particle occupation $n_f(n,s)$.  Also, in the bulk, due to
charge neutrality,  the Hartree potential cancels the background
potential and the energy of the single-particle HF level is
\begin{align}
E(n_f(m_1,s))=\sum_{m_2} [\delta_{m_1,m_2} m_1 \hbar \omega_c -E_{ex}(m_1,m_2)n_f(m_2,s)]
\end{align}
where the exchange energy is 
\begin{align}
E_{ex}(m_1,m_2)=\frac{1}{2} \int \frac{d^2q}{(2\pi)^2} v(\vec{q}) \rho_{m_1m_2}(\vec{q}) \rho_{m_2m_1}(-\vec{q}).
\end{align}
The matrix elements of the density operator are, for $n_1>n_2$,
\begin{align}
\rho_{n_1n_2}(\vec{q}) &= \sqrt{\frac{n_2!}{n_1!}} \bigg(\frac{q\ell \exp^{-i \theta_q}}{\sqrt{2}}\bigg)^{(n_1-n_2)} \nonumber \\ & L_{n_2}^{(n_1-n_2)}\bigg(\frac{q^2\ell^2}{2}\bigg) e^{-\frac{q^2\ell^2}{4}}
\end{align}
where $L_{n}^{m}$ is the associated Laguerre polynomial. For $n_1<n_2$
we use $\rho_{n_2n_1}(\vec{q})=(\rho_{n_1n_2}(-\vec{q}))^*$. Given the
occupations, the bulk ground state energy (at zero Zeeman energy) is
\begin{align}
  &{\cal{E}}_{gs} = N_{\phi} ( \sum\limits_{m,s} m\hbar\omega_c n_f(m,s) \nonumber \\ \quad\quad
  &-\frac{1}{2}\sum\limits_{m_1,m_2,s}E_{ex}(m_1,m_2)n_f(m_1,s)n_f(m_2,s))
\end{align}
where $N_\phi=\frac{eBL_xL_y}{h}$ is the number of flux quanta penetrating the sample.

In our work we use a screened Coulomb interaction, with
$v(q)=2\pi\hbar\omega_c{\tilde E}_c /(q+q_0)$, with $q_0$ being a screening
wavevector.  The unscreened Coulomb interaction corresponds to
$q_0=0$. Using the expression above one can show that for
$q_0=0,\tilde E_z=0$, the bulk $\nu=3$ state undergoes a phase transition
from a partially polarized state $(0\uparrow,0\downarrow,1\uparrow)$
to a fully polarized state $(0\uparrow,1\uparrow,2\uparrow$) at 
${\tilde E_c} =2.52$, whereas for the bulk $\nu=4$
state, the phase transition from an unpolarized state
$(0\uparrow,0\downarrow,1\uparrow,1\downarrow)$ to a fully polarized
state $(0\uparrow,1\uparrow,2\uparrow,3\uparrow)$ occurs at ${\tilde
  E_c} =2.90$. Hence there exists a regime of ${\tilde E_c}$ between
2.52 to 2.90, where the $\nu=3$ bulk phase is fully polarized and the
$\nu=4$ bulk phase is unpolarized.

\begin{figure}[H]
\includegraphics[width=0.45\textwidth]{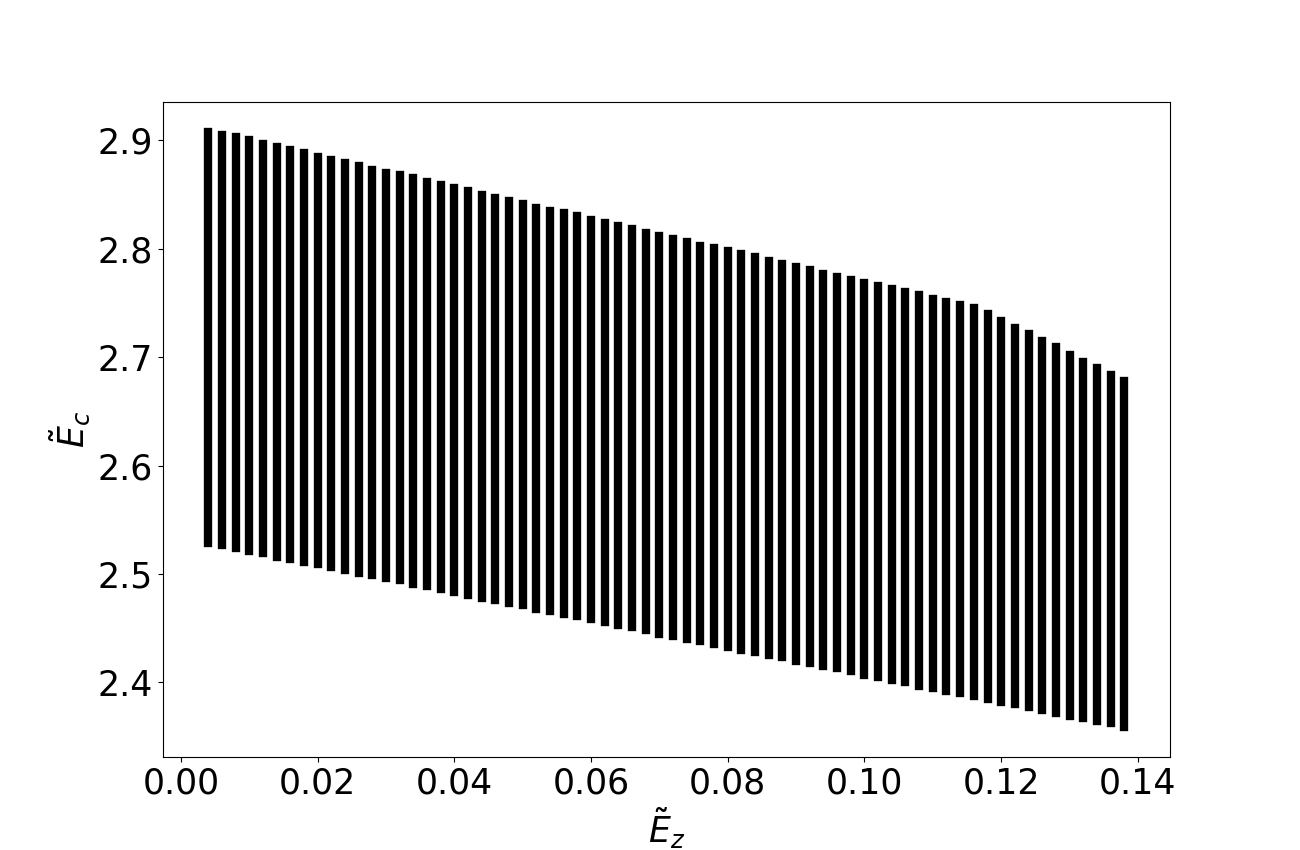}
\caption{The shaded region shows where the $\nu=4$ bulk is a singlet and the $\nu=3$ bulk is fully 
polarized for $q_0=0.01$. As ${\tilde E_z}$ increases, polarized states become lower in energy, 
and both boundaries shift to lower values of ${\tilde E_c}$. }
\label{fig:bulkphaseEz0}
\end{figure}

Note that changing the screening length and/or $\tilde E_z = E_z/\hbar\omega_c$ will shift both
phase transitions slightly, but there  remains a robust range of
${\tilde E_c}$ where the $\nu=3$ bulk phase is fully polarized and the
$\nu=4$ bulk phase is unpolarized.  The region in parameter space
where $\nu=4$ is a singlet and $\nu=3$ is fully polarized for $q_0=0.01$
is shown in Fig.\ref{fig:bulkphaseEz0}.

We will be working in the shaded region of the above phase
diagram where the structure of the interface between the bulk $\nu=4$
and bulk $\nu=3$ promises to be interesting - there is a change in
spin by 3/2 and the occupied Landau levels also change.

Now we go to our system of interest, with an interface between the
bulk $\nu=4$ and $\nu=3$ regions.  Once the HF averages are taken, the
HF Hamiltonian becomes diagonal in the guiding centre labels $k$. The
sample is assumed to be an infinite cylinder, with the translation
invariant $x$-direction taken to be periodic with a finite
circumference of  $20\pi l$, implying that our system has 10
guiding centers per magnetic length $l$.  We truncate our Hilbert
space so that it consists only of 4 Landau levels for each spin. We
then define an `active' region with a size of $50 l$ around the origin
whose $\Delta$-matrix is allowed to vary in the HF procedure, and a
`frozen' region on either side of the active region of size $45 l$,
whose occupations are fixed, respectively, to be those of the bulk
states $\nu=4$ and $\nu=3$ to the left and right of the active
region. The 'frozen' region simulates the Hartree and Fock
contributions of the bulk. The self-consistent ground state is then
found following a standard iterative procedure, described in detail in
Ref.\cite{Khanna2017}.  We have checked that changing the size of the
active and frozen regions does not change the results.

\section{II.\,\,\,\,\,\,   The phase diagram with three Landau levels}

In principle, the self-consistent HF should be carried out including
Landau level mixing to all Landau levels. This being
computationally impossible, one is forced to truncate the Hilbert
space by eliminating the Landau levels beyond some cutoff. In the main
paper we have quoted our HF results keeping four Landau levels in the
calculation.  We find two distinct edge phases - phase A obtained for
smooth edges or large values of ${\tilde w} =w/l$ with three
spin-resolved chiral modes and phase B obtained for sharp edges with a
single chiral mode with spin rotation, and obtained the phase diagram
of the two phases in the ${\tilde E_c}-{\tilde w}$ plane.

In this section, we present the phase diagram when only three Landau
levels are kept in the HF calculation (which is the minimum needed to
accomodate the fully polarized $\nu=3$ state). This allows us to
``switch off'' some of the Landau-level mixing, and allows us to infer
what would occur if we kept even more Landau levels than the four we
keep.

\begin{figure}[H]
\includegraphics[width=0.45\textwidth]{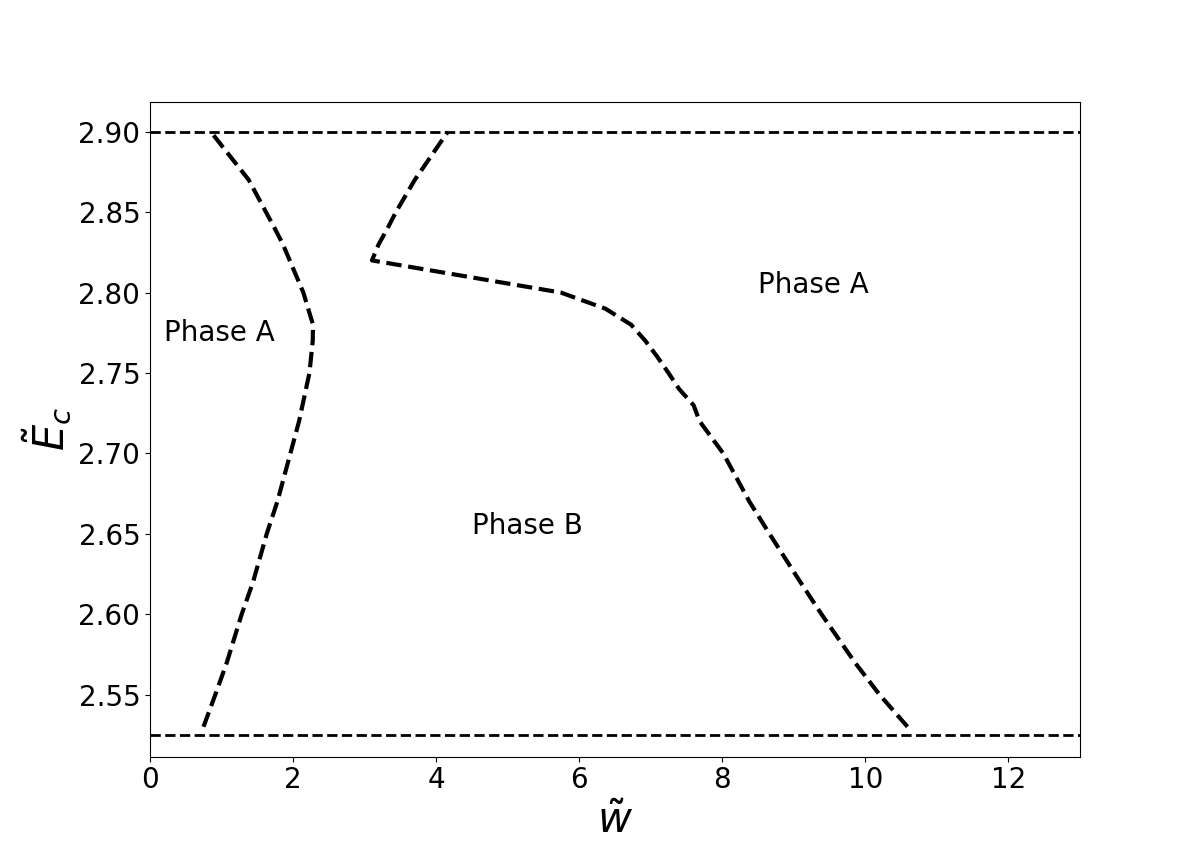}
\caption{Phase diagram in the parameter space of ${\tilde E_c}$ and
  ${\tilde w}$ at $\tilde E_z=0.002$ when only three Landau levels are
  kept in the HF calculation. The horizontal lines demarcate the upper
  and lower bounds of ${\tilde E}_c$ between which bulk $\nu=4$ is a
  singlet and bulk $\nu=3$ is fully polarized}
\label{fig:bulkphase3LL}
\end{figure}

As we see in Fig.\ref{fig:bulkphase3LL}, while the phase diagram for
${\tilde w}>3$ is quite similar to that with four Landau levels kept,
there is a difference at smaller ${\tilde w}$, with phase A being
re-entrant at most values of ${\tilde E}_c$.  However, the difference
between the ground state energies of the two phases is extremely tiny
in this region of the phase diagram. Increasing Landau level mixing
has the primary effect of reducing the spin-stiffness, which lowers
the energy of phase B without much altering that of phase A. Hence, it is
plausible that including more Landau levels in the calculation will
only further favor phase B, and that the phase diagram with four
Landau levels kept is qualitatively correct.

We have checked the stability of the phase diagram with four Landau
levels kept by increasing $\tilde E_z$ to 0.2 and we found phase A to
be absent for small values of $\tilde w$. In the following section we
will see in more detail the effect of $\tilde E_z$ and $q_0$ on our
phase diagram.

\section{III.\,\,\,\,\,\,  Robustness with respect to screening and the Zeeman coupling}

In this section, we  study the stability of the phase diagram
when the various parameters in the theory are changed.  Our main aim
here is to show that the HF phase diagram presented in the main text
is robust to variations of interaction parameters and $E_z$ within
reasonable, physically relevant limits.

\subsection{A :
Variation with respect to the Zeeman energy ${\tilde E_z}$}

In $GaAs$, in a magnetic field purely perpendicular to the sample, the
physical value of the Zeeman coupling is ${\tilde E}_z\approx0.03$. In
the main text, all the plots have been shown for $\tilde E_z = 0.03$.

Here, we restrict ourselves to keeping three Landau levels in the HF
calculation, which eases the computational burden enough to allow us
to go to low values of ${\tilde E_z}$ for all ${\tilde w}$.  We will
concentrate on the part of the phase diagram with ${\tilde w}>3$,
which as we have seen in the previous section, is qualitatively
identical to the phase diagram obtained by keeping four Landau
levels. 
We note that increasing
$\tilde E_z$ robustly favors phase A for smaller $\tilde E_c$, but
narrowly favors phase B for larger $\tilde E_c$.

\begin{figure}[H]
\includegraphics[width=0.45\textwidth]{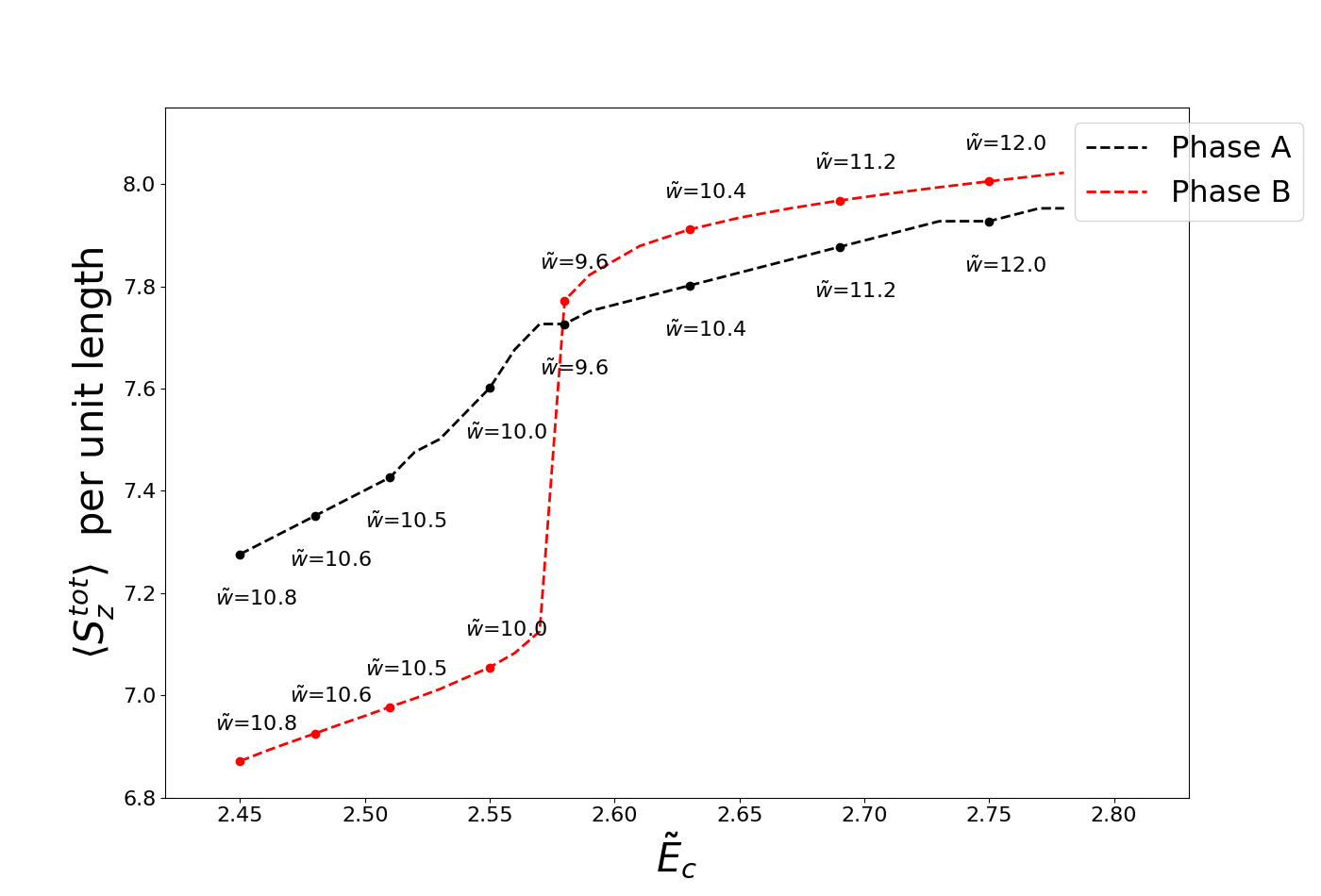}
\caption{Magnetization versus $\tilde E_c$ for $q_0=0.01$ and $\tilde E_z=0.08$. The ${\tilde w}$ 
values at each ${\tilde E}_c$ are chosen to be at the phase boundary. }
\label{fig:magnetization}
\end{figure}

From the Feynman-Hellman theorem, it is clear that
\begin{align}
  \frac{1}{N_\phi}\frac{\partial {\cal E}_{gs}}{\partial E_z}=-\langle HF| S^{tot}_z|HF\rangle
\end{align}
where the average on the right is in the HF ground state. Clearly, the
state with the larger average $S_z$ is favored upon increasing
$E_z$.  In Fig.\ref{fig:magnetization} we can see that for ${\tilde
  E}_c<2.57$, the total spin polarization of phase B is smaller than
that of phase A, and thus phase A expands at the expense of phase B as
$E_z$ increases. The opposite is true for ${\tilde E}_c>2.57$. Since
the difference in spin polarizations is smaller for ${\tilde E}_c>2.57$, the shift of the
phase boundary is smaller as well. It should be noted that most of the
increased spin polarization for larger ${\tilde E}_c$ occurs in the
$\nu=4$ region as a result of increased Landau level mixing.

\begin{figure}[H]
\includegraphics[width=0.45\textwidth]{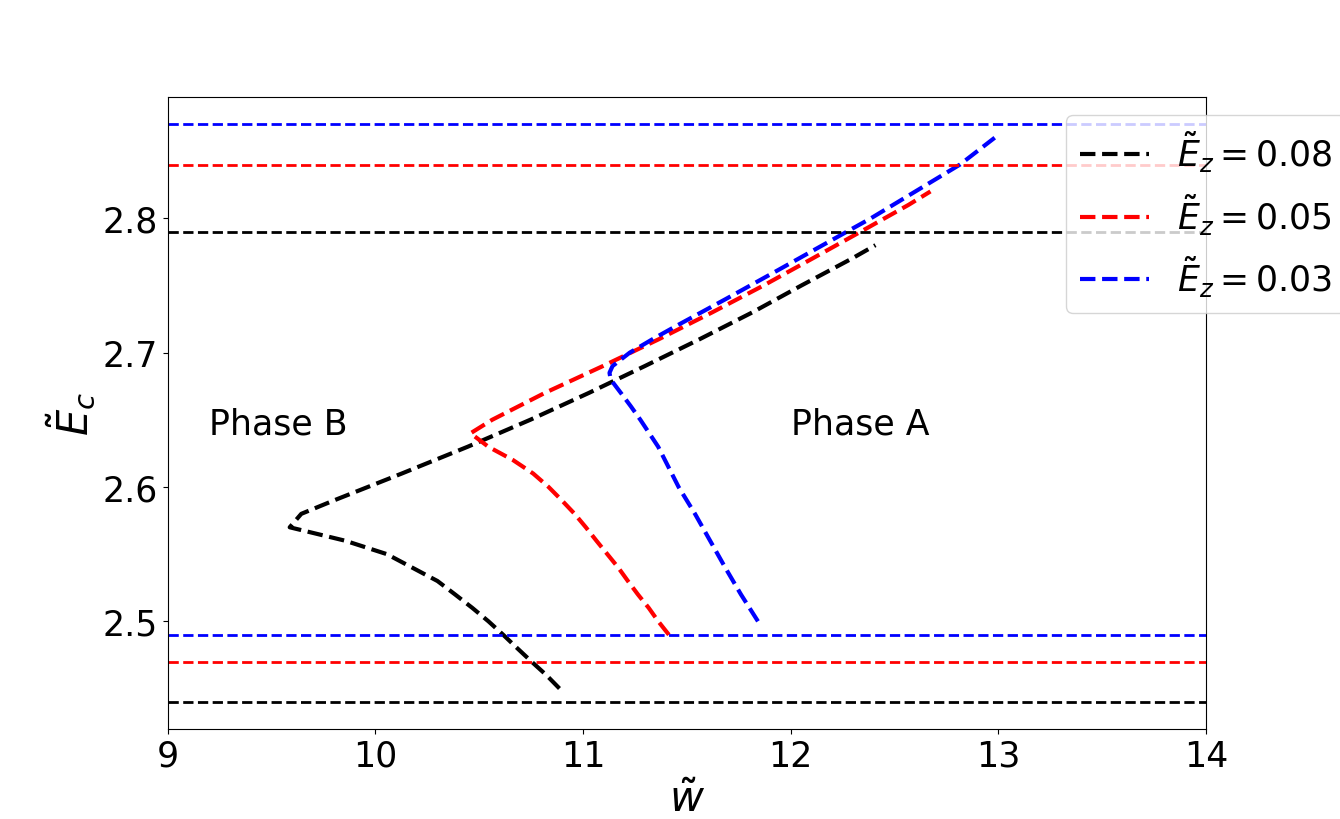}
\caption{Phase diagram for different values of ${\tilde E}_z$. As earlier, the
  coloured horizontal lines demarcate the upper and lower bounds of
  ${\tilde E}_c$ between which bulk $\nu=4$ is a singlet and bulk
  $\nu=3$ is fully polarized for the corresponding values of ${\tilde
    E}_z$.}
\label{fig:vzeeman}
\end{figure}

\subsection{B: Variation with respect to the screening length}

\begin{figure}[H]
\includegraphics[width=0.45\textwidth]{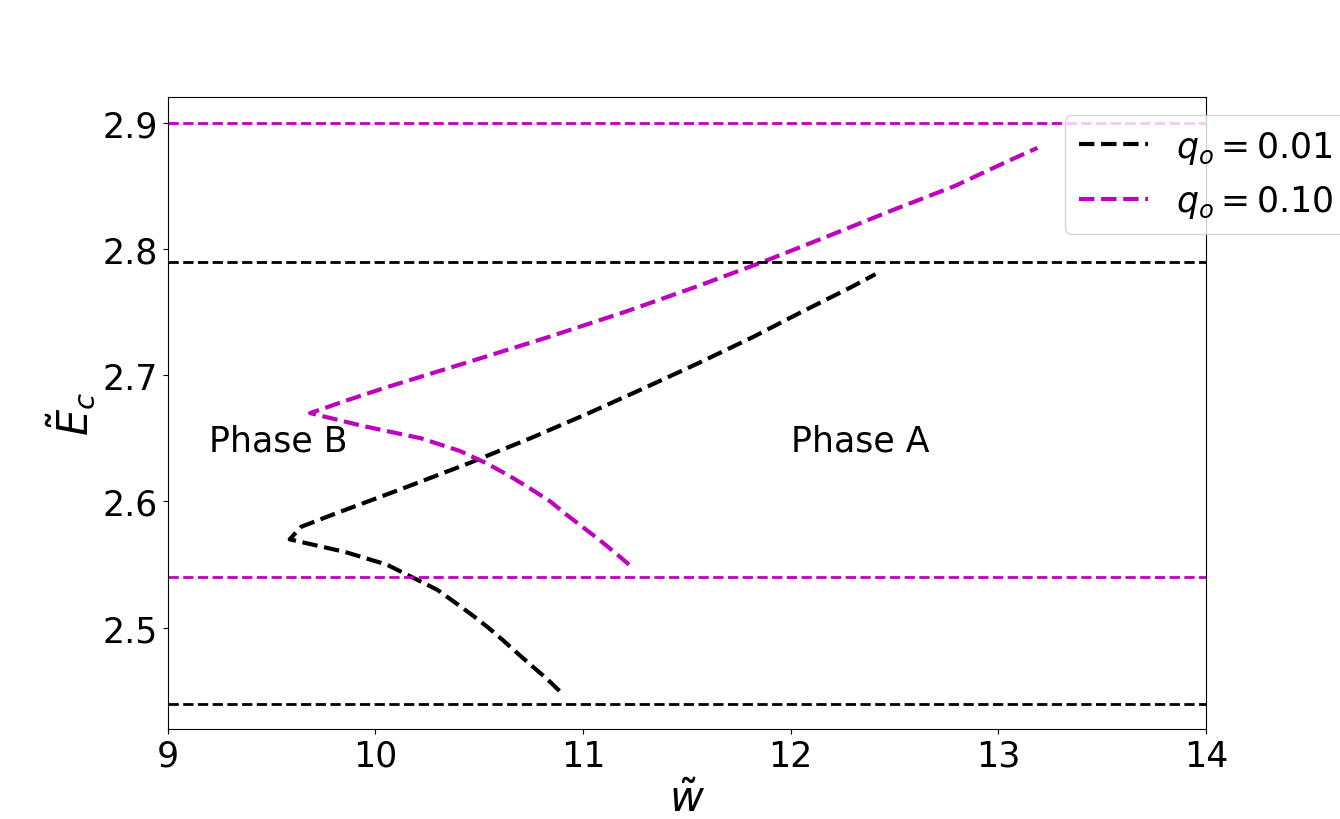}
\caption{Phase diagram for different values of the screening length $q_0$ at $\tilde E_z=0.08$.  Here again, the coloured horizontal lines 
demarcate the upper and lower bounds of
  ${\tilde E}_c$ between which bulk $\nu=4$ is a singlet and bulk
  $\nu=3$ is fully polarized for different $q_0$ values}
\label{fig:vq0}
\end{figure}

The parameter $q_0$ is a measure of the inverse screening length. In
our main paper we have shown the plots for $q_0=0.01$. In
Fig.\ref{fig:vq0} we show a comparison of the phase diagram between
$q_0=0.01$ and $q_0=0.1$ which has a smaller screening length as compared to
$q_0=0.01$. As can be seen, the phase diagrams are qualitatively
similar, with some quantitative shifts. We thus conclude that the
phase diagram is robust to changes in the screening length.

\subsection{C: Variation with respect to change in ratio of Hartree and exchange terms}

We have also varied the ratio of the Hartree and Fock terms to find if
Phase A(B) is favored by Hartree or the exchange terms. While this
does not correspond to any physical interaction, it can be a useful
way to gauge the importance of direct versus exchange terms in the two
phases.

We kept the strength of the Fock term at unity, in order to keep the
region of ${\tilde E}_c$ over which the desired bulk states are realized,
fixed.  The Hartree term was allowed to vary between 0.6 and 1.4. We
found that a Hartree term above unit strength favors phase A (the
region of phase A expands at the expense of phase B), while a Hartree term below
unit strength favors phase B.

  

\end{document}